\documentclass[twocolumn,showpacs,preprintnumbers,amssymb]{revtex4}

\usepackage{graphicx}			
\usepackage{bm}				

\newcommand{\expect}[1]{\left\langle#1\right\rangle}
\newcommand{\of}[1]{\left(#1\right)}
\newcommand{\abs}[1]{\left|#1\right|}
\newcommand{\qvec}{\mathbf{q}}
\newcommand{\rvec}{\mathbf{r}}
\newcommand{\xvec}{\mathbf{x}}

\newcommand{\kvec}{\mathbf{k}}
\newcommand{\Qvec}{\mathbf{Q}}
\newcommand{\Rvec}{\mathbf{R}}
\newcommand{\anticomm}[2]{\left\{#1,#2\right\}}
\newcommand{\comm}[2]{\left[#1,#2\right]}
\newcommand{\bea}{\begin{eqnarray}}
\newcommand{\eea}{\end{eqnarray}}
\newcommand{\beq}{\begin{equation}}
\newcommand{\eeq}{\end{equation}}
\newcommand{\bra}[1]{\left\langle#1\right|}
\newcommand{\ket}[1]{\left|#1\right\rangle}
\newcommand{\sgn}[1]{\frac{#1}{\abs{#1}}}
\newcommand{\barefreq}{\mu}
\newcommand{\ec}[1]{c_{#1}^{\dagger}}
\newcommand{\ed}[1]{c_{#1}^{\,}}
\newcommand{\qc}[1]{\psi_{#1}^{\dagger}}
\newcommand{\qd}[1]{\psi_{#1}^{\,}}
\newcommand{\nns}[1]{\eta_{#1}^*}
\newcommand{\nn}[1]{\eta_{#1}^{\,}}

\newcommand{\nuhat}{\hat{\bm{\nu}}}

\begin{document}

\preprint{APS/123-QED}

\title{Three-dimensional flux states as a model for the pseudogap 
phase of transition metal oxides}
\author{D. F. Schroeter}
\email{dfs@Stanford.edu}
\author{S. Doniach}
\affiliation{Department of Physics, Stanford University, Stanford, CA  94305}

\date{\today}

\begin{abstract}
We propose that the pseudogap state observed in the transition metal
oxides can be explained by a three-dimensional flux state, which
exhibits spontaneously generated currents in its ground state due to
electron-electron correlations.  We compare the energy of the flux
state to other classes of mean field states, and find that it is
stabilized over a wide range of $t$ and $\delta$.  The signature of
the state will be peaks in the neutron diffraction spectra, the
location and intensity of which are presented. The dependence of the
pseudogap in the optical conductivity is calculated based on the
parameters in the model.
\end{abstract}

\pacs{71.10.Fd, 71.30.+h, 75.10.Lp}
\keywords{SrRuO$_3$, pseudogap, flux state}
\maketitle

The motivation for this work is the observation of a pseudogap that
opens up in optical conductivity measurements of the three-dimensional
transition metal oxide SrRuO$_3$~\cite{geballeprl} above its
ferromagnetic transition temperature of T$_C \approx 150$ K. A
pseudogap has also recently been seen in BaRuO$_3$~\cite{baro3}. In
this pseudogapped regime, $\rho\of{T}$ increases linearly with
temperature, passing through the Ioffe-Regel limit without
saturation~\cite{geballejpc}, behavior indicative of a ``bad
metal''~\cite{badmetal}.  The optical conductivity in this state is
proportional to the non-Fermi liquid behavior of $\omega^{-1/2}$ at
high frequency and has a peak at low frequencies~\cite{geballeprl} at
approximately 250 cm$^{-1}$, the precise location of the peak being
temperature dependent.

We propose that this pseudogap state can be understood by considering
a ground state with spontaneously generated electronic currents
circulating around the plaquettes.  The currents arise from
electron-electron correlations, due to the bi-quadratic terms in the
Hamiltonian. The state which we propose is a generalization of the
two-dimensional flux states invented by Affleck and
Marston~\cite{a&m}, and studied in their chiral extension by Wen,
Wilzcek, and Zee~\cite{ww&z}.  Unlike the two-dimensional case, there
is no possibility of fractional statistics in three dimensions.
However, the spontaneous generation of gauge fields is a possibility
in three dimensions, and these gauge fields can lead to a ground state
with circulating electronic currents. Earlier work was done on
three-dimensional flux states by Laughlin and
coworkers~\cite{laughlin3d1, laughlin3d2} and Zee~\cite{zee1}.

In actuality, SrRuO$_3$ has 5 bands crossing the Fermi surface formed
by hybridizing the Ruthenium d orbitals with the Oxygen p
orbitals~\cite{singh}.  The crystal structure is orthorhombic,
becoming cubic at temperatures greater than 900 K~\cite{chak}.
Undoubtedly, the actual electronic structure of SrRuO$_3$, and
particularly the presence of a van Hove singularity near the Fermi
surface, influence the material's behavior.  The model which we
consider is vastly simplified and serves as a starting point for
considering the nature of the pseudogapped state in the
three-dimensional transition metal oxides.  A model which incorporates
some of these electronic features, but does not focus on the pseudogap
regime, has been proposed by Laad and
M{\"{u}}ller-Hartmann~\cite{muller}.

\section{Model System}
\label{sec:meanfield}

The Hamiltonian that we consider is the single orbital $t$-$J$
model given by
\beq
	H = J \sum_{\expect{i j}} \mathbf{S}_{i} \cdot \mathbf{S}_{j}
	- t \sum_{\expect{i j} \sigma} \ec{i \sigma} \ed{j \sigma} ,
	\label{eq:firsteq}
\eeq
where the sum over $\expect{i j}$ denotes nearest neighbors on a cubic
three-dimensional lattice.  Implicit in this equation is that we have
set $U = \infty$. The hopping matrix element $t$ which appears in
Equation~\ref{eq:firsteq} is taken to be an effective hopping element,
which has been greatly reduced due to this on-site Coulomb repulsion.
The value of $t$ will be set by two calculations in this paper: the
stability of the flux phase versus other mean field states calculated
in Section~\ref{sec:phase}, and the value of the optical conductivity
calculated in Section~\ref{sec:optical}.

The spin operators may be written in terms of the fermion operators to
give the Hamiltonian
\beq
	H = - \frac{J}{2} \sum_{\expect{i j}} \sum_{\sigma \sigma'}
	\ec{{i \sigma}} \ed{j \sigma} \ec{j \sigma'} \ed{i \sigma'} -
	t \sum_{\expect{i j}} \sum_{\sigma} \ec{i \sigma}
	\ed{j \sigma} + \frac{3 \mathcal{N} J}{4} .
\eeq
The Hamiltonian will be treated in the mean field, or Hartree-Fock
approximation. We make the replacement:
\beq
	\sum_{\sigma} \ec{i \sigma} \ed{j \sigma} \rightarrow
	\chi_{i j} + \of{ \sum_{\sigma} \ec{i \sigma} \ed{j \sigma} -
	\chi_{i j}}
\eeq
The assumption is made that the term in brackets, which corresponds to
fluctuations about the mean-field $\chi_{i j}$, is small and can be
included only to linear order. The resulting Hamiltonian is given by
\beq
	H = \frac{J}{2} \sum_{\expect{i j}} \left\{ \abs{\chi_{i j}}^2
	- 2 \left[ \of{\chi_{i j} + \frac{t}{J}} \ec{j} \ed{i} +
	\mathrm{H.c.} \right]\right\} , \label{eq:model}
\eeq
where we have dropped the redundant spin index. There is no a priori
reason to believe that the fluctuations about the mean field will be
small, although it has been rigorously shown in the two-dimensional
case for the large $n$ limit~\cite{a&m}, where $n$ is the particle
spin.

\begin{figure}
	\includegraphics{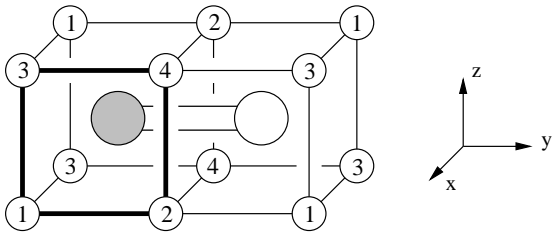} \caption{Unit cell for
	symmetry-broken bond Hamiltonian.  The dark line passes
	through the four atoms in the planar unit cell.  The spheres
	at the center of the cubes represent oppositely charged Dirac
	monopoles with the ``tail'' running through the interface
	between the two cubes.} \label{fig:unitcell}
\end{figure}

We allow the $\chi_{i j}$ to break the translational symmetry of the
lattice.  We choose a four-atom unit cell as shown in
Figure~\ref{fig:unitcell}. The lattice is generated by the primitive
translation vectors $\of{1,0,1}$, $\of{1,0,-1}$, and $\of{0,2,0}$, in
units of the bondlength $b$. The $\chi_{i j}$ in
Equation~\ref{eq:model} are then parameterized by 12 complex numbers.
We use the notation $\chi_{i \nu}$, where the index $i = 1, \ldots, 4$
is the location of the atom in the unit cell and the index $\nu$ gives
the direction $\left\{x,y,z\right\}$.

This choice is made because it allows for the formation of a $\pi$-per
plaquette flux phase, something that a two atom unit cell does not
allow in three dimensions.  The model is computationally simpler than
the eight-band model studied by Laughlin~\cite{laughlin3d1,
laughlin3d2}, at the expense of picking out a preferred direction.  To
get a feel for what this corresponds to, one can think of the gauge
fields in the sample as being generated by Dirac monopoles of
alternating charge sitting at the centers of each cube.  The ``tails''
of the monopoles are connected to form dipoles.  The $\hat{\bm{y}}$
direction in our model corresponds to the dipolar axis.

The self-consistency of the model is the requirement that the energy
as determined by Equation~\ref{eq:model} be a local minimum with
respect to variations of the twelve complex parameters $\chi_{i \nu}$.
This can be seen by writing
\begin{eqnarray}
	\expect{H} & = & \frac{\mathcal{N} J}{8} \sum_{i = 1}^4
	\sum_{\nu = 1}^3 \left\{ \abs{\chi_{i \nu}}^2
	\right. \nonumber \\ & & \left. - 2 \left[\of{\chi_{i \nu}^* +
	\frac{t}{J}} \expect{\ec{i} \ed{i + \nu}} + \mathrm{H.c.} 
	\right]\right\} .
\end{eqnarray}
Minimizing this function with respect to $\chi_{i \nu}^*$ gives the
self-consistency relation
\beq
	\expect{\ec{i} \ed{i + \nu}} = \frac{\chi_{i \nu}}{2} .
\eeq
Note also that at this minimum, the expectation value of the
Hamiltonian is given by
\beq
	\expect{H} = - \frac{\mathcal{N} J}{8} \sum_{i = 1}^4
	\sum_{\nu = 1}^3 \left[\abs{\chi_{i \nu}}^2 + 2 \frac{t}{J} \,
	\mathrm{Re}\of{\chi_{i \nu}}\right] .  \label{eq:atminimum}
\eeq

The Hamiltonian in Equation~\ref{eq:model} contains both a field
strength and a term which we define as the bond Hamiltonian:
\beq
	H_B = J \sum_{\expect{i j}} \sum_{\sigma} \of{\chi_{j i} +
	\frac{t}{J}} \ec{i \sigma} \ed{j \sigma} + \mathrm{H.c.}
\eeq
The bond Hamiltonian can be diagonalized by introducing a set of
operators
\beq
	\psi^{\dagger}_{\qvec \lambda} = \frac{1}{\sqrt{\mathcal{N}}}
	\sum_{\ell} e^{i \qvec \cdot \rvec_{\ell}} u_{\qvec
	\lambda}\of{\ell} c_{\rvec_{\ell}}^{\dagger} ,
	\label{eq:quasiparticle}
\eeq
where $\mathcal{N}$ is the number of sites in the lattice, $\lambda$
is the band index which runs from $1$ to $4$ and the $u_{\qvec
\lambda}\of{\ell}$ are a set of functions periodic in the unit cell. 
The bands are determined by the eigenvalue equation
\beq
	H_{\qvec} u_{\qvec \lambda} = \epsilon_{\qvec \lambda}
	u_{\qvec \lambda} ,
\eeq
where $H_{\qvec}$ is written in terms of the hopping elements and
$u_{\qvec \lambda}$ is a 4 component vector $u_{\qvec \lambda}\of{1}
\cdots u_{\qvec \lambda}\of{4}$.  For the unit cell depicted in 
Figure~\ref{fig:unitcell}, the Hamiltonian takes the form
\beq
	H_{\mathbf{q}} = J \of{\begin{array}{cccc} 0 & \nn{1} & \nn{3}
	& 0 \\ \nns{1} & 0 & 0 & \nn{4} \\ \nns{3} & 0 & 0 & \nn{2} \\
	0 & \nns{4} & \nns{2} & 0 \end{array}} , \label{eq:hamq}
\eeq
where
\begin{eqnarray}
	\nn{1} & = & \tilde{\chi}_{1y} e^{i q_y} + \tilde{\chi}_{2y}^*
	e^{-i q_y} \nonumber \\ \nn{2} & = & \tilde{\chi}_{3y} e^{i
	q_y} + \tilde{\chi}_{4y}^* e^{-i q_y} \nonumber \\ \nn{3} & =
	& \tilde{\chi}_{1x} e^{i q_x} + \tilde{\chi}_{1z} e^{i q_z} +
	\tilde{\chi}_{3x}^* e^{-i q_x} + \tilde{\chi}_{3z}^* e^{-i
	q_z} \nonumber \\ \nn{4} & = & \tilde{\chi}_{2x} e^{i q_x} +
	\tilde{\chi}_{2z} e^{i q_z} + \tilde{\chi}_{4x}^* e^{-i q_x} +
	\tilde{\chi}_{4z}^* e^{-i q_z} .
\end{eqnarray}
The tilde in the expression above means that these numbers include the
actual hopping element: $\tilde{\chi}_{i \nu} = \chi_{i \nu} +
t/J$. The eigenenergies for the four bands are given by
\beq
	\epsilon_{\mathbf{q}} = \pm J \sqrt{ \frac{\nn{i} \nns{i}}{2}
	\pm \sqrt{\of{ \frac{\nn{i} \nns{i}}{2}}^2 - \abs{\nn{1}
	\nns{2} - \nn{3} \nns{4}}^2}} , \label{eq:geneq}
\eeq
with the usual Einstein summation convention.  In
Section~\ref{sec:flux} we will consider the band structure and
eigenstates of one particular solution of the bond Hamiltonian, the
flux state where the hopping elements are complex and equal in
magnitude: $\abs{\chi_{i \nu}} = \chi$ for all $i,\nu$.

For $t = 0$ the gauge fields $\chi_{i j}$ are unobservable.  However,
as soon as $t$ is increased they lead to real circulating electronic
currents.  Consider the site $i$ and the six sites $j$ which are its
nearest neighbors. The sum of the currents flowing outward from site
$i$ is then the time rate of change of the number operator $n_i$ on
site $i$
\bea
	\sum_{j} j_{i j} & = & \expect{\frac{\partial}{\partial t}
	n_{i}} = \frac{i}{\hbar} \expect{\comm{H_B}{n_{i}}} ,
	\nonumber \\ & = & \frac{2 i}{\hbar} \sum_{j} \of{J \chi_{i j}
	+ t} \expect{\ec{j} \ed{i}} - \mathrm{H.c.} ,
\eea
where the factor of 2 arises from the spin. The term $\chi_{i j}
\expect{\ec{j} \ed{i}}$ is real and gives no 
contribution, but for $t > 0$ and $\chi_{i j}$ complex, the bonds
carry electronic currents.  These currents are the signature of the
flux state.  They have long-ange order and can be probed by neutron
diffraction as will be shown in Section~\ref{sec:neutron}.

\section{The Phase diagram}
\label{sec:phase}

The Hamiltonian described in Section~\ref{sec:meanfield} admits a
number of self-consistent solutions.  We have performed a numerical
search as a function of the hopping element $t$ and the doping
$\delta$.  The function to be minimized is
\beq
	E = \frac{\mathcal{N}}{8} \left[J \sum_{i \nu} \abs{\chi_{i
	\nu}}^2 + 2 \frac{1}{\mathcal{N}_m} \sum_{\kvec \lambda}'
	\epsilon_{\lambda \kvec} \right] \label{eq:minimizefunction} ,
\eeq
where $\mathcal{N}_m = \mathcal{N}/4$ is the number of unit cells in
the lattice.  The search is performed using Powell's method in the
space of the $\chi_{i \nu}$.  At each point in the space, the bands
are determined by Equation~\ref{eq:geneq}, and the lower magnetic
bands are filled up with $N_e = \mathcal{N} \of{1 - \delta}$
electrons.

Our search is limited to three classes of states.  We consider the
flux phase with $\abs{\chi_i} = \abs{\chi}$ for all $i$ with the
phases of the hopping elements unconstrained.  At $t = \delta = 0$ the
flux phase has flux $\Phi = \pi$ per plaquette as defined by
\begin{equation}
	e^{i \Phi} = \prod_{\mathrm{plaquette}} \frac{\tilde{\chi}_{i
	\nu}}{\abs{\tilde{\chi}_{i \nu}}} .  \label{eq:fluxdef}
\end{equation}
Note that a flux of $\pi$ and a flux of $-\pi$ per plaquette are
indistinguishable since in either case the electron acquires a phase
of $-1$ upon traversing the plaquette.  This is the chiral symmetry of
the model.  Away from $t=\delta=0$, the flux per plaquette is no
longer $\pi$ and the chiral symmetry is broken.

Another state that we have considered is the dimer state.  In this
case each site forms a bond with a neighboring site.  In the case of
$t = \delta = 0$, one particular manifestation of the state is
$\chi_{1z} = \chi_{4x} = 1$ with all other $\chi_{i \nu} = 0$.  The
dimer state has flat bands $\epsilon_{\qvec} = \pm J$ and can be
considered to be a charge-density state with the charge localized on
the bonds for which $\chi_{i \nu} = 1$~\cite{a&m}. Away from the point
$t = \delta = 0$, the state becomes partially dimerized. There are in
principle a number of non-equivalent dimer solutions, but the solution
mentioned above appears to be the lowest energy configuration.

At values of $\delta \gtrsim 0.2$ another local minimum is the kite
state which has $\chi_{1y} = \chi_{2x} = \chi_{3z} = \chi_{4y} \neq
\chi_{i \nu}$.  The state is called a kite state because the lines of 
charge, if the analogy from the dimer state above is used, form
zigzagging patterns through the lattice, and could lead to a lattice
distortion from the large Coulomb repulsion between charged
lines. This particular instantiation of the kite state is chosen for
the same reasons as the dimer above. The last state which is
considered is the uniform state with all $\chi_{i \nu}$ equal and real
valued.  This state is a simple Fermi liquid, with a renormalized
value of the hopping matrix element.

\begin{figure}
	\includegraphics{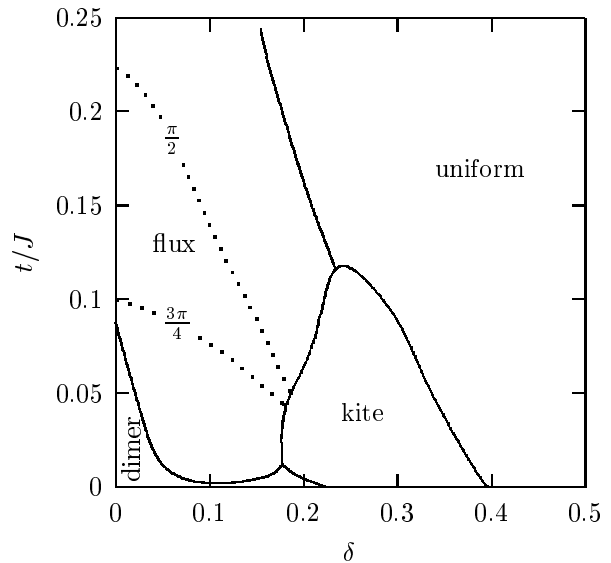} \caption{Phase Diagram for Bond States.
	The diagram shows the competition between the four types of
	states considered in the text.  The contours in the flux phase
	region show the average flux per plaquette, as defined by
	Equation~\ref{eq:fluxdef}.} \label{fig:phasediag}
\end{figure}

The phase diagram is shown in Figure~\ref{fig:phasediag}. While we
have performed the calculation for an arbitrary flux state as
described above, the phase diagram shows a calculation performed using
a restricted parameter set.  The parameters used are shown as
different arrows in Figure~\ref{fig:numcube} below. This is done
because at $\delta \approx 0.1$, there is a transition to a flux phase
with a different type of ordering which is outside the scope of the
current discussion.  Apart from this transition, there is no
qualitative change in the phase diagram when the unconstrained flux
state is considered. From the phase diagram we see that the flux state
is stabilized over a fairly wide range of doping, but that the flux
per plaquette decreases from the value of $\pi$ as one leaves the
point $t = \delta = 0$.  In the calculations which follow we will
assume that $t/J = 0.1$, a point at which the highly symmetric
$\pi$-per plaquette flux state discussed below in
Section~\ref{sec:flux} is a reasonable approximation to the actual
mean field state.

One must also consider the possibilities of other types of order which
are not described by the mean-field $\chi_{i j}$. The most insidious
of these is antiferromagnetic order.  At $t = \delta = 0$, the
energies of the two stabilized states discussed above are
\beq
	E_{\mathrm{dimer}} = - \frac{\mathcal{N} J}{4}, \, \, \, \, \,
	\, E_{\mathrm{flux}} \approx -0.95 \, \frac{\mathcal{N} J}{4}
	.
\eeq
For comparison, the N\'{e}el state, which is characterized by
\begin{equation}
	\chi_{i j} = 0 \, \, \, \, \, \, \, \, \, \expect{\ec{i
	\sigma} \sigma^z \ed{i \sigma'}} = \of{-1}^i,
\end{equation}
has an energy of $- 3 \mathcal{N} J / 4$, lower than either of the two
bond states at $t = \delta = 0$.  In order for the bond states to be
actualized, a term has to be added which will destabilize the
antiferromagnetic order.  This can be done by adding a next-nearest
neighbor hopping term $J'$~\cite{doniach}, in which case the energy of
the N\'{e}el state will be
\beq
	E_{\hbox{\scriptsize{N\'eel}}} = - \frac{3 \mathcal{N} J}{4}
	\of{1 - 2 \frac{J'}{J}} .
\eeq
The energies of the bond states are actually unchanged up to a value
of $J' / J \approx 1/3$, which is the threshold for acquiring a
nonzero value of the next nearest neighbor $\chi_{i j}$ as shown by
Laughlin and Zou~\cite{laughlinzou}.  At the value $J' = J / 3$, the
energy of the N\'{e}el state is equal to the energy of the dimer state
and very close to the energy of the flux state.  Doping will also
serve to destabilize the antiferromagnet so that the crossover will
actually occur at a lower value of $J'$ than the one reported
here. Therefore, while it is not treated explicitly in this paper,
some term such as the next-nearest neighbor $J'$ must be added to this
model in order to make the bond states energetically favorable with
respect to the N\'{e}el ordered state.

\section{The Flux Phase}
\label{sec:flux}

\begin{figure}
	\includegraphics{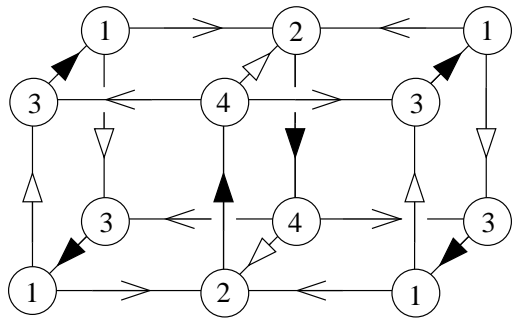} \caption{The $\pi$ per plaquette
	flux phase. All bonds have the same magnitude. The arrows
	correspond to complex phases of $\phi$ ($>$) , $\phi - \pi/4$
	($\vartriangleright$), and $3 \pi / 4 - \phi$
	($\blacktriangleright$), where $\phi$ is determined by $\tan
	\phi = \sqrt{2}$.  This diagram shows the same portion of the
	lattice as is shown in Figure~\ref{fig:unitcell}.} 
	\label{fig:numcube}
\end{figure}

In the calculations which follow, we consider the $\pi$ per plaquette
flux phase described in Section~\ref{sec:phase} which is only truly
stabilized at $\delta = t = 0$.  Some care must be used in selecting
this state since at $t = \delta = 0$ the system is invariant under a
gauge transformation where
\bea
	\ed{i} \rightarrow e^{i \phi_{i}} \ed{i}, \, \, \, \, \, \,
	\chi_{i j} \rightarrow e^{i \of{\phi_{j} - \phi_{i}}} \chi_{i
	j} .
\eea
Away from this point, this symmetry disappears, and the low-energy
state is the one for which the quantity
\beq
	\sum_{i \nu} \mathrm{Re} \left[\chi_{i \nu}\right] ,
	\label{eq:realcriteria}
\eeq
is a maximum as can be seen from Equation~\ref{eq:atminimum}.  The
state is shown in Figure~\ref{fig:numcube}.  It can be found by either
maximizing the function in Equation~\ref{eq:realcriteria} for an
arbitrary gauge transformation, or by numerically continuing a state
from $t > 0$ down to $t = 0$.  The choice of the correct
symmetry-breaking gauge is important since it will affect the
distribution of electronic currents in the sample and hence observable
features such as the neutron diffraction and optical conductivity.

\begin{figure}
	\centerline{\includegraphics{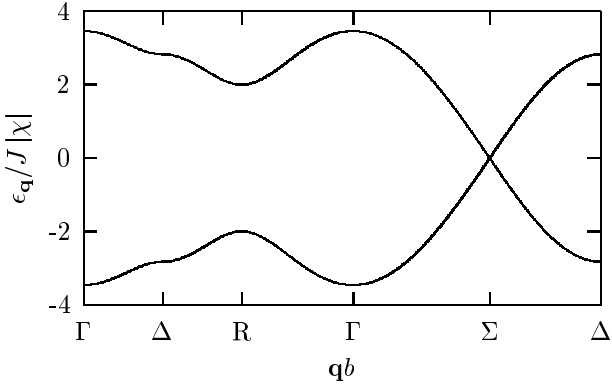}} \caption{Flux Phase band
	structure. Points in the Brillouin zone are $\Gamma =
	\of{0,0,0}$, $\Delta = \of{\pi/2,0,0}$, $R = \of{\pi/2, 0,
	\pi/2}$, and $\Sigma = \of{\pi/2, \pi/2, \pi/2}$. There are
	two distinct Dirac points located at $\qvec b = \of{\pi/2,
	\pi/2,\pi/2}$ and $\qvec b = \of{\pi/2, \pi/2,-\pi/2}$.} 
	\label{fig:bstruct}
\end{figure}

In this state, the Hamiltonian in Equation~\ref{eq:quasiparticle} can
be rewritten using a set of Dirac matrices $\alpha_x$, $\alpha_y$, and
$\alpha_z$ such that
\beq
	H_{\qvec} = 2 J \abs{\chi} \sum_{\nu} \cos\of{q_{\nu} b}
	\alpha_{\nu} ,
	\label{eq:hamdirac}
\eeq
where the matrices satisfy the algebra
\beq
	\anticomm{\alpha_i}{\alpha_j} = 2 \delta_{i j} .
	\label{eq:diracalgebra}
\eeq
Explicit forms for the matrices are given in
Appendix~\ref{sec:appendix}. It is found numerically that $\abs{\chi}
\approx 0.4$. The eigenvalues of the Hamiltonian are twofold
degenerate and are given by
\beq
	\epsilon_{\qvec} = \pm 2 J \abs{\chi} \sqrt{\sum_{\nu}
	\cos^2\of{q_{\nu} b}} .
\eeq
The band structure is shown in Figure~\ref{fig:bstruct}.  At
half-filling the Fermi surface reduces to two isolated points at
$\qvec b = \of{\pi/2, \pi /2, \pi / 2}$ and $\qvec b = \of{\pi/2,
\pi/2, -\pi/2}$ shown at the point $\Sigma$ in Figure~\ref{fig:bstruct}.  
The low energy excitations about these points are relativistic.

In order to calculate the neutron cross-section and the optical
conductivity, we also will need the eigenvectors which appear in
Equation~\ref{eq:quasiparticle}.  The matrix $\mathbf{u}$ whose rows
correspond to the bands $1 \ldots 4$ and whose columns correspond to
the position in the unit cell is given by
\beq
	\mathbf{u}_{\qvec} = \frac{1}{\sqrt{2}}
	\of{\begin{array}{cccc} 
	\abs{\eta_{\qvec}} & - \sgn{\eta_{\qvec}} e^{- i \phi} & 0 &
	\sgn{\eta_{\qvec}} \gamma_{\qvec} \\
	\abs{\gamma_{\qvec}} & 0 & \sgn{\gamma_{\qvec}} e^{-i \phi} &
	- \eta_{\qvec} \sgn{\gamma_{\qvec}} \\
	\abs{\eta_{\qvec}} & \sgn{\eta_{\qvec}} e^{- i \phi} & 0 &
	-\gamma_{\qvec} \sgn{\eta_{\qvec}} \\
	\abs{\gamma_{\qvec}} & 0 & - \sgn{\gamma_{\qvec}} e^{- i \phi}
	& - \eta_{\qvec} \sgn{\gamma_{\qvec}} \end{array}} ,
	\label{eq:umat}
\eeq
where we have defined the quantities
\begin{eqnarray}
	\eta_{\qvec} = \frac{\cos q_y b}{\Gamma_{\qvec}} \, \, \, \,
	\, \, \, \, \, & \gamma_{\qvec} & = e^{-i \frac{\pi}{4}}
	\frac{\cos q_x b - i \cos q_z b}{\Gamma_{\qvec}} \nonumber \\
	\Gamma_{\qvec} & = & \sqrt{\sum_{\nu} \cos^2\of{q_{\nu} b}} .
	\label{eq:defs}
\end{eqnarray}
The phases of the eigenvectors in Equation~\ref{eq:umat} have been
selected so that the eigenstates in Equation~\ref{eq:quasiparticle} are
invariant under $\qvec \rightarrow \qvec + \Qvec$ where $\Qvec$ is any
vector in the reciprocal lattice.

\section{Neutron Scattering}
\label{sec:neutron}

The flux states can be probed by neutron scattering, as the neutron
spin interacts with the magnetic dipoles generated by the real
electron currents circulating on the plaquettes. The interaction
potential~\cite{affleckneutron} is written
\beq
	V\of{\rvec} = 2 \sum_{\expect{i j}} t \of{\ec{i} \ed{j} -
	\mathrm{H.c.}} \exp\left[\frac{i e}{\hbar c}
	\int_{\xvec_i}^{\xvec_j} \mathbf{A} \cdot d\ell \right] ,
\eeq
where the vector potential is given by
\beq
	\mathbf{A} = \bm{\mu} \times \frac{\rvec_e -
	\rvec_n}{\abs{\rvec_e - \rvec_n}^3}, \, \, \, \, \, \, \, \,
	\bm{\mu} = - \gamma \frac{e \hbar}{m_n c} \mathbf{S} .
\eeq
In these expressions $\mathbf{S}$ is the neutron spin and $\gamma
\approx 1.91$ is a constant. It can be shown~\footnote{ The
reasoning leading to this result is identical to that of Affleck, Hsu
and Marston for the two-dimensional flux
states~\cite{affleckneutron}.} that
\bea
	\int d\rvec e^{i \qvec \cdot \rvec_n} V\of{\rvec} & = & i
	\of{\gamma r_0} \of{\frac{m}{m^*}} \frac{8 \pi \hbar^2}{m_n}
	\nonumber \\ & & \times \frac{1}{\abs{\qvec b}^2} \sum_{\nu}
	J_{\nu} \frac{\nuhat \cdot \of{\mathbf{S} \times
	\qvec}}{\qvec \cdot \nuhat} , \label{eq:potlint}
\eea
with the current operator defined as
\bea
	J_{\nu} & = & \sum_{\kvec} \ec{\kvec + \qvec} \ed{\kvec}
	f_{\nu}\of{\kvec,\qvec} \nonumber \\ f_{\nu}\of{\kvec, \qvec}
	& = & \cos\of{k_{\nu} b} - \cos\of{k_{\nu} b + q_{\nu} b} .
\eea
In these expressions we have replaced the hopping element $t$ by
$\hbar^2 / 2 m^* b^2$, with $m^*$ the effective mass. Since we are
assuming (see Section~\ref{sec:phase}) that $t/J = 0.1$, the effective
mass will be quite large.  If we assume that $J$ takes the typical
value of 0.1 eV, we have $m^* \approx 15 \, m$. The vector $\nuhat$ is
a unit vector with the sum running over the $x$, $y$, and $z$
directions.

Converting this to a cross-section and averaging over the spin,
assuming that the incoming beam is unpolarized, the expression for the
cross-section is given by
\beq
	\frac{d \sigma}{d\Omega} = \of{\gamma r_0}^2
	\of{\frac{m}{m^*}}^2 \frac{4}{\abs{\qvec b}^4} \sum_{\nu <
	\nu'} \abs{\frac{q_{\nu}}{q_{\nu'}} \expect{J_{\nu'}} -
	\frac{q_{\nu'}}{q_{\nu}} \expect{J_{\nu}}}^2 .
	\label{eq:cross}
\eeq
In order to calculate the matrix elements in Equation~\ref{eq:cross},
we need to rewrite the current operator in terms of the eigenstates of
our system. We first break up the momentum sum so that it runs only
over the first Brillouin zone
\beq
	J_{\nu} = \sum_{\kvec}' \sum_{i = 1}^4 \ec{\kvec + \qvec +
	\Qvec_i} \ed{\kvec + \Qvec_i} f_{\nu}\of{\kvec + \Qvec_i,
	\qvec} , \label{eq:neutronfirst}
\eeq
where $\Qvec_1 b = \mathbf{0}$, $\Qvec_2 b = \of{\pi, 0, \pi}$,
$\Qvec_3 b = \of{0, \pi, 0}$, and $\Qvec_4 b = \bm{\pi}$.  We can
rewrite the electron operators at momentum $\kvec + \Qvec$ in terms of
the $c_{\qvec}\of{\ell}$ defined as
\beq
	c_{\qvec}^{\dagger}\of{\ell} = \frac{1}{\sqrt{\mathcal{N}_m}}
	\sum_{\Rvec} e^{i \qvec \cdot \of{\Rvec + \rvec_{\ell}}}
	c_{\Rvec + \rvec_{\ell}}^{\dagger} ,
\eeq  
where $\Rvec$ runs over all the unit cells and $\rvec_{\ell}$ is the
position of the $\ell^{\mathrm{th}}$ atom in the unit cell. This
introduces a matrix $g$ with elements $g_{i \ell} = \exp\left[i
\Qvec_i \cdot \rvec_{\ell}\right]/2$, and results in the current operator
\beq
	J_{\nu} = \sum_{\kvec}' \sum_{i m p} g_{i m} g_{i p}
	\ec{\kvec + \qvec}\of{m} \ed{\kvec}\of{p}
	f_{\nu}\of{\kvec + \Qvec_i, \qvec} .
\eeq
We also note that $f_{\nu}\of{\kvec + \Qvec_i, \qvec} =
\left[\bar{Q}^{\nu}\right]_{ii} f_{\nu}\of{\kvec,\qvec}$, where the 
matrix $\bar{Q}^{\nu}$ is diagonal with elements $\exp\left[ i \Qvec_i
\cdot \bm{\nu} b\right]$. The sum on $i$ can then be performed to obtain 
\beq
	J_{\nu} = \sum_{\kvec}' \sum_{m p} \ec{\kvec + \qvec}\of{m}
	\left[g \bar{Q}^{\nu} g\right]_{m p} \ed{\kvec}\of{p}
	f_{\nu}\of{\kvec, \qvec} .  \label{eq:metricintroduce}
\eeq
The matrix $g \bar{Q}^{\nu} g$ in the above expression has a natural
interpretation.  It merely connects all sites in the lattice which are
connected by a hopping element in the $\bm{\nu}$ direction.  It can be
written in the form
\beq
	g \bar{Q}^{\nu} g = \of{\begin{array}{cc} \sigma^x & 0 \\
	0 & \sigma^x \end{array}} \delta_{\nu y} +
	\of{\begin{array}{cc} 0 & \openone \\ \openone & 0
	\end{array}} \of{\delta_{\nu x} + \delta_{\nu z}} .
	\label{eq:metric}
\eeq
Finally, we can rewrite Equation~\ref{eq:metricintroduce} by inverting
the eigenvector matrix in Equation~\ref{eq:umat}. This gives
\beq
	J_{\nu} = \sum_{\kvec}' \sum_{\lambda \lambda'}
	\left[u_{\kvec} \of{g \bar{Q}^{\nu} g} u_{\kvec +
	\qvec}^{\dagger}\right]_{\lambda' \lambda} f_{\nu}\of{\kvec,
	\qvec} \qc{\lambda, \kvec + \qvec} \qd{\lambda', \kvec} .
	\label{eq:jfinalneutron}
\eeq
If we assume that the lower bands are completely filled, we can then
write a simple expression for the expectation values of these matrix
elements.
\beq
	\expect{J_{\nu}} = \sum_{\bm{\tau}} \sum_{\kvec}' \,
	\mathrm{Tr}' \left[u_{\kvec} \of{g \bar{Q}^{\nu} g} u_{\kvec +
	\qvec}^{\dagger} \right] f_{\nu}\of{\kvec, \qvec}
	\delta_{\qvec \bm{\tau}} \label{eq:tracematrix}
\eeq
The prime on the trace indicates that it runs only over the lower two
bands, and the sum on $\bm{\tau}$ runs over all vectors in the
reciprocal lattice.

Expression~\ref{eq:tracematrix} is quite general and can be used to
calculate properties away from zero doping by restricting the sum on
$\kvec$ such that $\kvec < \kvec_F$.  Additionally, this equation
assumes nothing about the actual structure of the eigenvector matrix
$u_{\kvec}$. Specializing to the case of the flux state discussed in
Section~\ref{sec:flux}, the trace for $\nu = y$ is
\beq
	i \of{1 - \frac{\gamma_{\kvec + \qvec}^*}{\gamma_{\kvec}^*}}
	\eta_{\kvec} \sin \phi .
\eeq
Here we have used the fact that $\eta_{\kvec + \qvec} = -
\eta_{\kvec}$, a condition enforced by the presence of $f_{y}$ in 
Equation~\ref{eq:tracematrix}. We see that the trace is zero unless
$\gamma_{\kvec + \qvec} = - \gamma_{\kvec}$, which means that
$\expect{J_y}$ vanishes unless $q_{\nu} b$ is an odd multiple of $\pi$
for all $\nu$.  For the $\nu = x$ and $\nu = z$ cases, the traces are
the same as can be seen from Equation~\ref{eq:metric}.  They are given
by
\beq
	i \of{1 - \frac{\eta_{\kvec + \qvec}}{\eta_{\kvec}}}
	\mathrm{Im} \gamma_{\kvec} \cos \phi - i \of{1 +
	\frac{\eta_{\kvec + \qvec}}{\eta_{\kvec}}} \mathrm{Re}
	\gamma_{\kvec} \sin \phi ,
\eeq
where we have taken $\gamma_{\kvec + \qvec} = - \gamma_{\kvec}$ for
the same reasons as above.  This expression takes two different values
depending on whether $q_y b$ is an even or an odd multiple of $\pi$.

\begin{figure}
	\includegraphics{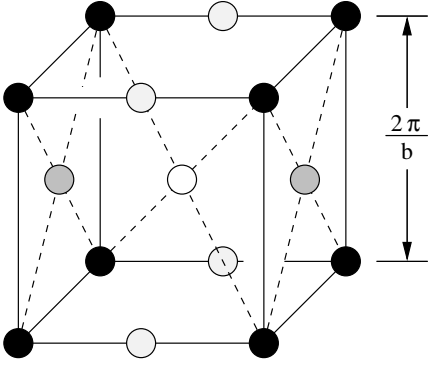} \caption{Reciprocal Lattice
	Vectors for Flux Phase.  The black dots show the scattering
	from the nuclear centers.  The white dot shows the
	antiferromagnetic scattering vector at $\qvec b =
	\bm{\pi}$. The shaded dots at the face centers show those
	points in the reciprocal lattice space of the flux phase which
	produce scattering, and the lighter shaded dots along the line
	centers show those points in the reciprocal lattice space
	which do not produce scattering.}  \label{fig:reciprocate}
\end{figure}

Scattering will only occur at the reciprocal lattice vectors, as
guaranteed by the delta function in Equation~\ref{eq:tracematrix}. The
reciprocal lattice is shown in Figure~\ref{fig:reciprocate}.  There is
no magnetic scattering at the nuclear locations, since they occur at
even multiples of $\pi$.  There is additionally no scattering at the
line centers $\qvec b = \pi \hat{\bm{y}}$, since all three matrix elements
vanish at these points. Scattering does occur at both the face-centers
at $\qvec b = \of{\pi,0,\pi}$ and at the body-center at $\qvec b =
\bm{\pi}$.  The former distinguishes the scattering from Bragg
scattering from a cubic antiferromagnet. It is likely that the actual
material would consist of domains containing all dipolar orientations,
so that scattering would actually be observed at all the face-centers.

The magnitude of the scattering will in general be quite small.  We
write the cross section as
\beq
	\frac{d \sigma}{d \Omega} = \mathcal{N}_m \frac{\of{2
	\pi}^3}{v_{0 m}} \of{\gamma r_0}^2 \of{\frac{m}{m^*}}^2
	\sum_{\bm{\tau}} M\of{\qvec} \delta\of{\qvec - \bm{\tau}} ,
	\label{eq:bigscatter}
\eeq
where we have rewritten the Kronecker delta function from
Equation~\ref{eq:tracematrix} in terms of the Dirac delta function
with the proper normalization of $\of{2 \pi}^3 / V$. The term $v_{0m}
= 4 b^3$ is the volume of the unit cell.  The structure factor is
given by
\begin{widetext}
\beq
	M\of{\qvec} = \frac{4}{3} \frac{\abs{\chi}^2 }{\abs{\qvec
	b}^2} \left[ \of{3 + \cos q_y b} \of{\frac{1}{\of{q_x b}^2} +
	\frac{1}{\of{q_z b}^2}} + 4 \frac{1 - \cos q_y b}{\of{q_y
	b}^2} \right] \of{1 - \cos q_x b} \of{1 - \cos q_z b} .
	\label{eq:neutronform}
\eeq
\end{widetext}
In deriving this expression, we have repeatedly used the fact that
\beq
	\frac{1}{\mathcal{N}_m} \sum_{\kvec}' \frac{\cos k_{\nu} \cos
	k_{\nu'}}{\Gamma_{\kvec}} = \abs{\chi} \delta_{\nu \nu'} ,
\eeq
with the $\Gamma_{\kvec}$ defined as in Equation~\ref{eq:defs}. This
relation follows from the symmetry of the momentum sum and
Equation~\ref{eq:minimizefunction}. The structure factor in
Equation~\ref{eq:neutronform} takes the same value at the two smallest
scattering angles corresponding to the points $\Qvec_2 b = \of{\pi, 0,
\pi}$ and $\Qvec_4 b = \bm{\pi}$ in the reciprocal lattice:
\beq
	M\of{\qvec} = \frac{64}{3 \pi^4} \abs{\chi}^2
\eeq
To give a feel for the order of magnitude of the scattering from the
flux states, we compare to the scattering from an antiferromagnet.  In
that case one has
\beq
	\left. \frac{d \sigma}{d \Omega} \right|_{AF} = \frac{2}{3}
	\mathcal{N}_m^{AF} \frac{\of{2 \pi}^3}{v_{0m}^{AF}} \of{\gamma
	r_0}^2 \expect{S^{\eta}}^2 \sum_{\bm{\tau}}
	\abs{F\of{\bm{\tau}}}^2 \delta\of{\qvec - \bm{\tau}} ,
\eeq
where the factor of $2/3$ arises by assuming that the sublattice
magnetization is along a crystallographic axis.  We can estimate the
form factor for the antiferromagnet by assuming it is the same as that
of chromium, a typical band antiferromagnet.  Chromium has a form
factor of $F\of{\bm{\pi}/b} \approx 0.4$~\cite{fawcett}. The unit cell
in the flux phase is twice as large as the antiferromagnetic unit cell
and therefore $\mathcal{N}_m^{AF} = 2 \mathcal{N}_m$ and $v_{0m}^{AF}
= v_{0m}/2$.  Assuming that the spins are 50\% polarized such that
$\expect{S^{\eta}} = 1/4$, and taking $m^* = 15 m$, we see that the
scattering from the antiferromagnet is roughly $170$ times larger than
the scattering from the flux state at the wave-vector $\qvec b =
\bm{\pi}$.  Note that the discrepancy in size is due primarily to the
size of the effective hopping element $t$, which must be small
compared to $J$ if the flux phase is going to be stabilized away from
$\delta = 0$ as was shown in Section~\ref{sec:phase}.

\section{Optical Conductivity}
\label{sec:optical}

In our model, the peak in the optical conductivity arises from
transitions between the bands shown in Figure~\ref{fig:bstruct}.  The
model is too simple to accurately predict the optical conductivity of
a material such as SrRu0$_3$. However, the calculation illustrates
both the dependence of the location of the peak on the spin-exchange
energy $J$ and the intensity of the peak on the ratio of $t/J$.

The calculation of the optical conductivity is very similar to that of
the neutron diffraction. In this case we couple the system to a
time-dependent vector potential $\mathbf{A} = A\of{t} \nuhat$, where
the vector potential's time dependence and relation to the electric
field are given by
\beq
	A\of{t} = A e^{-i \omega t} \,\,\,\,\,\,\,\,\,\, \mathbf{E} =
	\frac{i \omega}{c} \mathbf{A} .
	\label{eq:ave}
\eeq
We assume that the wavelength of the light is long enough that we can
ignore any spatial dependence in the fields. The vector potential
couples to the hopping terms in the Hamiltonian. A phase is acquired
according to
\beq
	\ec{j} \ed{i} \rightarrow \ec{j} \ed{i} \exp\left[\frac{i
	e}{\hbar c} \int_{\xvec_i}^{\xvec_j} \mathbf{A} \cdot
	d\bm{\ell}\right] .
\eeq
It is important to note that the corresponding $\chi_{i j}$
appearing in Equation~\ref{eq:model} also acquire an equal and
opposite phase so that when we expand the Hamiltonian to linear order
in the vector potential $A\of{t}$ we do not get any contribution from
the terms proportional to $\chi_{i j}$.  The perturbation to
the Hamiltonian is
\beq
	H' = - \frac{L}{c}  \mathbf{A} \cdot \mathbf{j} ,
	\label{eq:perturbed}
\eeq
where $L = \mathcal{N}^{1/3} b$ is the length of the sample and the
current operator in the $\bm{\nu}$ direction is given by
\beq
	j_{\nu} = \frac{2 i t e b}{\hbar L} \sum_{i} \left[\ec{i}
	\ed{i + \bm{\nu}} - \ec{i + \bm{\nu}} \ed{i}\right] .
	\label{eq:currentop}
\eeq
Note that the operator in Equation~\ref{eq:currentop} defines the
total current, not the current density, flowing in the $\bm{\nu}$
direction. The complex optical conductivity is related to the induced
current density in the sample by
\beq
	\mathbf{J}_{\mathrm{ind}} = \sigma \mathbf{E} .
	\label{eq:jve}
\eeq

The induced current can be calculated using linear response theory. In
that case it can be shown from Equations~\ref{eq:ave},
\ref{eq:perturbed} and \ref{eq:jve} that the real portion of the optical
conductivity is related to the complex portion of the current-current
correlation function 
\beq
	\sigma_1\of{\omega} = - \frac{1}{\omega} \mathrm{Im} \,
	\chi\of{\omega} ,
\eeq
where the current-current correlation function is given by
\bea
	\chi\of{\omega} & = & \frac{1}{L} \sum_n \abs{\bra{0} j
	\ket{n}}^2 \left[ \frac{1}{\hbar \omega - E_n + E_0 + i s}
	\right. \nonumber \\ & & - \left. \frac{1}{\hbar \omega + E_n
	- E_0 + i s}\right] .
\eea
The infinitesimal $s$ arises from assuming that the perturbation in
Equation~\ref{eq:perturbed} vanishes at $t=-\infty$.  The states
$\ket{0}$ and $\ket{n}$ are to be evaluated at $t = -\infty$ or
$\mathbf{A} = 0$.  The current operator in Equation~\ref{eq:currentop}
can be written in terms of the eigenstates of the system, following the
same approach applied in Equations~\ref{eq:neutronfirst}
through~\ref{eq:tracematrix} .  The result is that
\beq
	j_{\nu} = 4 \frac{t e b}{L \hbar} \sum_{\kvec}' \sum_{\lambda
	\lambda'} \left[u_{\kvec} \of{g \bar{Q}^{\nu} g}
	u_{\kvec}^{\dagger}\right]_{\lambda' \lambda} \sin\of{k_{\nu}
	b} \qc{\lambda \kvec} \qd{\lambda' \kvec} .  \label{eq:finalj}
\eeq
Equation~\ref{eq:finalj} is very similar to
Equation~\ref{eq:jfinalneutron}.  In this case, however, we are
considering matrix elements connecting the ground state to excited
states and so pick up the contributions at $\lambda \neq \lambda'$.
Averaging the optical conductivity over the three directions in the
lattice, it takes the form
\beq
	\sigma_1\of{\omega} = \frac{\pi}{12 \abs{\chi}^2}
	\frac{e^2}{\hbar b} \of{\frac{t}{J}}^2 \frac{1}{\barefreq}
	\frac{1}{\mathcal{N}_m} \sum_{\qvec}' F\of{\qvec}
	\delta\of{\barefreq - \Gamma_{\qvec}} , \label{eq:finalsigma}
\eeq
where $\Gamma_{\qvec}$ was defined in Equation~\ref{eq:defs} and where
we have defined a dimensionless frequency given by $\barefreq =
\hbar \omega / 4 J \abs{\chi}$.  The structure factor is given by:
\beq
	F\of{\qvec} = \sum_{\nu} \sum_{\lambda = 1}^{2} \sum_{\lambda'
	= 3}^4 \abs{\left[u_{\qvec} \of{g \bar{Q}^{\nu} g}
	u_{\qvec}^{\dagger}\right]_{\lambda' \lambda}}^2 \sin^2
	q_{\nu} b \label{eq:opticalstructure}
\eeq
The sums in Equation~\ref{eq:opticalstructure} can be performed to
give
\bea
	F\of{\qvec} & = & \of{\frac{4}{3} + \frac{\of{\gamma_{\kvec} -
	\gamma_{\kvec}^*}^2}{6}} \of{\sin^2 k_x b + \sin^2 k_z b}
	\nonumber \\ & & + \frac{2}{3} \of{1 + \eta_{\kvec}^2} \sin^2
	k_y b , \label{eq:lastf}
\eea
with the quantities defined as in Equation~\ref{eq:defs}.

\begin{figure}
	\centerline{\includegraphics{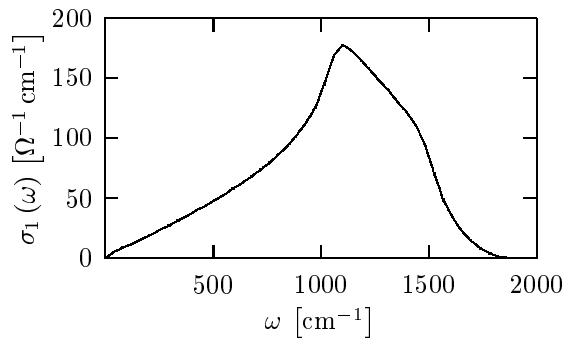}} \caption{Interband contribution
	to optical conductivity for the flux phase.}
	\label{fig:optical}
\end{figure}

Equations~\ref{eq:finalsigma} and~\ref{eq:lastf} have been evaluated
numerically, and the result is shown in Figure~\ref{fig:optical}.
Within our model, the location of the peak is proportional to the
exchange energy $J$.  Assuming this takes the value 0.1 eV, we find
that the peak occurs at approximately 1000 cm$^{-1}$, which gives
order of magnitude agreement with the observed value of $250$
cm$^{-1}$ in SrRuO$_3$~\cite{geballeprl}.  Due to the simplicity of
the model we are solving, one would not expect more accurate
agreement.  The magnitude of the peak is governed by the ratio of
$t/J$.  Direct comparison of this quantity with experiment is more
difficult. This is due to the fact that SrRuO$_3$ has five bands,
whereas we have considered only a single orbital.  Additionally, our
calculation only considers the interband contribution to the
conductivity, whereas the real material also has an intraband
contribution from thermally excited carriers.  If one assumes that the
result shown in Figure~\ref{fig:optical} needs to be scaled by a
factor of roughly five to account for the number of orbitals in
SrRuO$_3$, the results are reasonable compared with the measured
conductivity of 6000 $\Omega^{-1} \, \mathrm{cm}^{-1}$.

\section{Discussion}

We propose that the three-dimensional flux state is a good candidate
for the pseudogap state seen in the transition metal oxides.  Its
signature will be the presence of weak neutron diffraction peaks
arising from the ordered electronic currents in the material. The fact
that SrRuO$_3$ is not near an antiferromagnetic transition and the
fact that the three dimensional flux states produce scattering at
wavevectors other than $\qvec b = \bm{\pi}$ make the system an
excellent candidate in which to observe this type of order.  Further
theoretical work is warranted to understand how the actual electronic
structure of the material will influence the behaviors discussed here.

\begin{acknowledgments}
We would like to thank R. Krishna and J. Franklin for many useful
discussions and S. Kivelson for inspirational remarks.

We acknowledge support from the DOE through the Complex Materials
Program at SSRL.
\end{acknowledgments}

\appendix*

\section{}
\label{sec:appendix}

The matrices from Equations~\ref{eq:hamdirac}, \ref{eq:diracalgebra},
are given explicitly by
\bea
	\alpha_x & = & \of{\begin{array}{cccc} 0 & 0 & Z e^{- 3 i \pi
	/ 4} & 0 \\ 0 & 0 & 0 & Z^* e^{i \pi / 4} \\ Z^* e^{ 3 i \pi /
	4} & 0 & 0 & 0 \\ 0 & Z e^{- i \pi / 4} & 0 & 0
	\end{array}}\nonumber \\
	\alpha_z & = & \of{\begin{array}{cccc} 0 & 0 & Z e^{- i \pi /
	4} & 0 \\ 0 & 0 & 0 & Z^* e^{3 i \pi / 4} \\ Z^* e^{ i \pi /
	4} & 0 & 0 & 0 \\ 0 & Z e^{- 3 i \pi / 4} & 0 & 0 \end{array}}
	\nonumber \\ 
	\alpha_y & = & \of{\begin{array}{cccc} 0 & Z & 0 & 0 \\ Z^* &
	0 & 0 & 0 \\ 0 & 0 & 0 & Z^* \\ 0 & 0 & Z & 0 \end{array}}
	\label{eq:explicitmatrices} ,
\eea
where
\beq
 	Z = e^{i \phi} = \frac{1}{\sqrt{3}} + i \sqrt{\frac{2}{3}} .
\eeq

\end{document}